\documentclass[twocolumn]{aastex62}
\usepackage{lineno}

\usepackage{natbib}
\usepackage{float}
\graphicspath{{./}{figures/}}

\accepted{May 2025}
\submitjournal{ApJ}

\shorttitle{Tycho's SNR Ejecta Mass Ratios}
\shortauthors{Holland-Ashford et. al.}

\begin{document}

\title{Estimating Global Ejecta Mass Ratios in Tycho's SNR} 

\correspondingauthor{Tyler Holland-Ashford}
\email{tyler.e.holland-ashford@nasa.gov}
\author{Tyler Holland-Ashford}
\affil{Astrophysics Science Division, NASA Goddard Space Flight Center Greenbelt, MD 20771, USA}
\author{Patrick Slane}
\affil{Center for Astrophysics $|$ Harvard \& Smithsonian, 60 Garden St., Cambridge MA 02138, USA}
\author{Brian Williams}
\affil{Astrophysics Science Division, NASA Goddard Space Flight Center Greenbelt, MD 20771, USA}


\begin{abstract}
In this work, as a follow-up to our similar analysis of Kepler's supernova remnant (SNR), we estimate total mass ratios of various ejecta elements in Tycho's SNR using {\it Suzaku} X-ray data. In our spectral analysis, we account for uncertainties arising from {\it Suzaku}'s effective area calibration (5\%--15\%) and from the unknown filling factors of the various plasma components in our spectral model (1\%--10\%). We compare our calculated ejecta mass ratios to results from previous X-ray analyses of Tycho's SNR and to the nucleosynthesis results from Type Ia supernova simulations. Our estimated ejecta mass ratios for Tycho's SNR are only consistent with simulations that use a $\sim$90\% attenuated $^{12}$C$+^{16}$O reaction rate (as for Kepler's SNR), are inconsistent with simulations involving a double detonation of a thick helium layer, and support a Type Ia explosion of normal luminosity where $\sim$85\% of the ejecta has been heated by the reverse shock.

\end{abstract}


\section{Introduction}

The progenitor systems of many Type Ia supernovae (SNe)---the thermonuclear explosions of typically C$+$O white dwarfs (WDs)---remain uncertain. Most broadly, Type Ia SNe occur as a result of interactions between a primary WD and a secondary companion star. For single-degenerate (SD) explosions the companion is a main- or post-main-sequence star, while for double-degenerate (DD) explosions the companion is another WD \citep{hoyle60, colgate66, nomoto82}. However, the specific properties of both stars, the orbital dynamics of the system, and the type of accretion can all drastically affect explosion properties. For example, the central density, mass, and metallicity of the primary WD can significantly affect nucleosynthesis yields of intermediate-mass elements (IMEs; e.g., O, Mg, Ne, Si, S, Ar, Ca) and iron-group elements (IGEs; Cr, Mn, Fe, Ni) \citep{iwamoto99,bravo10,seitenzahl11}.

Additionally, the possibility of double detonations (DDets) of the primary can further complicate the SN Ia picture. In these scenarios, slow accretion onto the primary WD can result in a built-up layer of He that eventually ignites and prompts an inner core explosion \citep{taam80,livne90,alan19}. DDets can produce higher IGE/Fe ratios than single detonations of the same primary WD mass \citep{lach20}, and ones with unstable accretion (Dynamically-Driven Double-Degenerate Double Detonations: D$^6$) can result in an earlier explosion where the secondary WD may survive \citep{guillochon10,dan11,shen18b}.

Tycho's supernova remnant (SNR; SN 1572) is a young, $\sim$450~yr old remnant of Type Ia SNe \citep{krause08,rest08}. It is one of the few naked-eye Galactic SNe recorded in history \citep{stephenson02} and has distance estimates of 2.3--4~kpc \citep{chevalier80,fink94,schwarz95,williams13,kozlova18}. Its exact progenitor and explosion mechanisms are still not conclusively known. Asymmetry analysis by \cite{picquenot24} revealed an elliptic SNR where heavier elements exhibited more symmetric distributions than lighter elements, suggestive of a bipolar explosion with less turbulence in the innermost layers of the explosion. Additionally, \cite{zhou16} found evidence of an expanding molecular bubble around Tycho's SNR, and \cite{tanaka21,uchida24} found significant deceleration of the blast wave, both of which imply the existence of strong winds from a nondegenerate companion. However, a surviving companion has yet to be identified \citep{ihara07,kerzendorf09,xue15, kerzendorf18}.

In this paper, we calculate the mass ratios of various X-ray-emitting ejecta elements in Tycho's SNR in order to constrain its origin. As this paper is a follow-up to a previous paper on estimating ejecta mass ratios in Kepler's SNR (\citealt{ha23}, hereafter HA23), we follow much of the same methodology described therein. In Section~\ref{sec:methods}, we present the {\it Suzaku} observations used and the steps of our analysis, focusing on those that differ from HA23.  In Section~\ref{sec:results}, we present our final estimates for ejecta mass ratios observed in Tycho's SNR. In Sections~\ref{sec:pastwork} and \ref{sec:simuls}, we compare our results with the mass ratios observed in previous literature studies of Tycho's SNR and with nucleosynthesis yields predicted by various Type Ia simulations.

\begin{figure*}
\begin{center}
\includegraphics[width=0.93\textwidth]{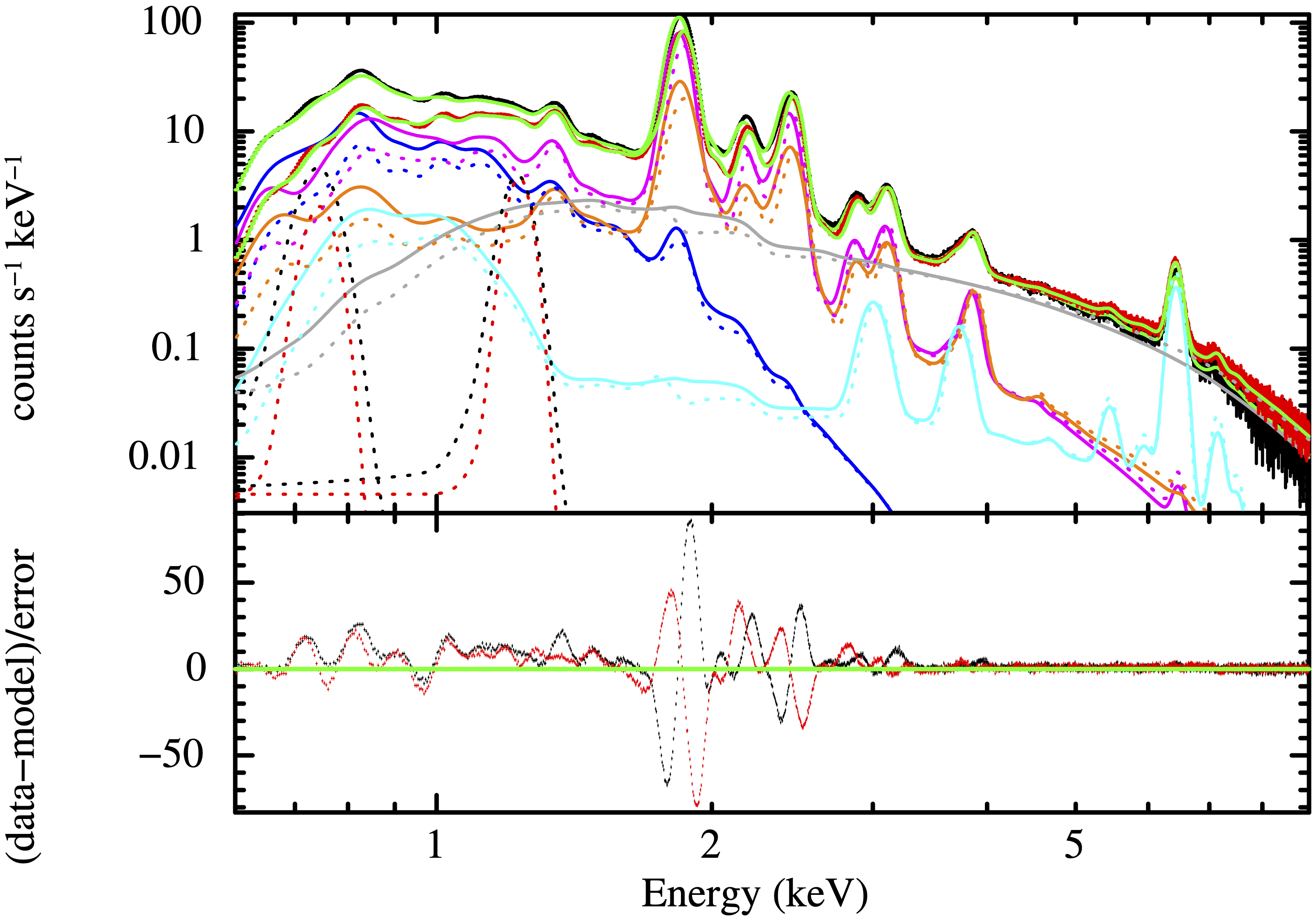}
\end{center}
\vspace{-5mm}
\caption{\footnotesize{{\it Suzaku} X-Ray spectra of Tycho's SNR and our best spectral fit. The black data is from the XIS1 detector and the red is from the combined XIS0$+$XIS3 detectors. The colored lines are components of our best-fit model: green is the full model; blue is shocked CSM/ISM; magenta and orange are the lower-temperature Si-dominated ejecta (ej1 \& ej2); cyan is the hotter Fe-dominated ejecta (ej3); and gray is synchrotron emission. The solid lines are fit to {\it Suzaku} XIS1 and the dotted lines are fit to the combined XIS0$+$XIS3 detectors. }}
\label{fig:specfit}
\end{figure*}

\section{Observations and Data Analysis}
\label{sec:methods}

\subsection{{\it Suzaku} Data Reduction}
Tycho's SNR was observed by {\it Suzaku} three times from 2006 to 2008 and, with an angular diameter of $\sim$8.5\arcmin\ \citep{green19}, entirely fit within the field of view of {\it Suzaku}'s X-Ray Imaging Spectrometer (XIS; \citealt{koyama07}) detectors. Following the {\it Suzaku} Data Reduction Guide\footnote{https://heasarc.gsfc.nasa.gov/docs/suzaku/analysis/abc/}, we reprocessed all three observations---ObsIDs 500024010, 503085020, and 503085010---using the HEADAS software version 6.33.2, the {\it Suzaku} calibration database (CALDB) files released in 2018-10-23, and the reprocessing tool \texttt{aepipeline}. We estimated the non-X-ray background (NXB) with \texttt{xisnxbgen} \citep{tawa08} and created the redistribution matrix file (rmf) and auxiliary response file (arf) with \texttt{xisrmfgen} and \texttt{xisarfgen}. Finally, we combined the data from the XIS0 and XIS3 (front-illuminated) detectors to fit simultaneously with the data from the XIS1 (back-illuminated) detector. 

\subsection{Spectral Fitting}
\label{sec:specanal}
We used XSPEC version 12.14.0h \citep{arnaud96} and AtomDB v3.0.10 \footnote{This version of AtomDB was a private version, obtained before the release of versions 3.1.0 through 3.1.2. It contains the updated inner shell transitions of highly ionized Fe listed in the release notes for v3.1.2. 
When we later fit Tycho's SNR using AtomDB v3.1.2, we obtained mass ratio estimates that were all within 1$\sigma$ of the ones found in this paper using AtomDB v3.0.10.}
\citep{smith01,foster12} to fit the 0.6--8.0~keV spectrum of Tycho's SNR. 

Our fitting process was essentially the same as for Kepler's SNR in \cite{ha23}. We used an absorbed multicomponent model, including one shocked plasma component to capture emission from swept-up circumstellar and/or interstellar medium (CSM, ISM), three shocked plasma components to capture ejecta emission at different states, a nonthermal component to capture synchrotron emission from the SNR's forward shock, and two Gaussians at $\sim$0.7 and $\sim$1.2~keV to account for residuals common to SNR plasma models---likely due to Fe-L lines, but the exact reason is uncertain (e.g., \citealt{okon20,uchida24}). These components were broadened to account for material moving at different velocities within the SNR, and each component's redshift term was left thawed to account for a bulk Doppler shift. In Xspec, our final model was \texttt{tbabs}*(\texttt{gsmooth}*\texttt{vpshock}$_{\rm CSM}$ + \texttt{gsmooth}*\texttt{vpshock}$_{\rm ej1}$ + \texttt{gsmooth}*\texttt{vpshock}$_{\rm ej2}$ + \texttt{gsmooth}*\texttt{vvnei}$_{\rm ej3}$ + \texttt{srcut} + \texttt{gaussian} + \texttt{gaussian}), where ej1 \& ej2 were IME dominated ([Si/H] fixed to 10$^5$ time solar), and ej3 was Fe-dominated ([Fe/H] fixed to 10$^5$ times solar). We froze the radio spectral index \texttt{srcut} to $\alpha=-0.6$ \citep{kothes06,tamagawa09,green19} and let the Ar and Ca abundances of ej3 vary freely. We use solar abundances from \cite{wilms00}. See HA23 a for further explanation of our model, and see Table~\ref{table:fitvals} in Appendix~\ref{appen:comprehensive} of this paper for our best-fit values and uncertainties.

An example of our spectral fit is shown in Figure~\ref{fig:specfit}. As shown, residuals are quite high (reduced-$\chi^2 > 5$), even more than for our analysis of Keplers SNR \citep{ha23}, and the fits to the front- versus back-illuminated detectors have differing---often opposite---residuals. We chose to use all the {\it Suzaku} data on Tycho's SNR rather than a subset, as we did for Kepler's SNR, because of these differing residuals. Using all the data presents the least unbiased spectrum of Tycho's SNR for fitting. Our usage of three detectors meant that we were analyzing about three times as much data of Tycho's SNR than we did for Kepler's SNR, which, combined with the strong detector-specific residuals not present in data of Kepler's SNR, explains why our fit to Tycho's SNR has so much worse residuals. However, we note that the Poisson photon errors are very small ($\lesssim$1\%) due to the extremely large number of counts from Tycho's SNR. In such a situation, it is critical to account for all other possible sources of error, including ones that are negligible when fitting much lower-quality spectra. In this case, we needed to include the systematic uncertainties of the {\it Suzaku} telescope that are not included in the X-ray data by default.

\subsection{Quantifying Uncertainties}

To fully account for telescope uncertainties, we follow the procedure described in HA23, which was in turn heavily adapted from \cite{lee11,xu14}, who accounted for telescope effective area uncertainties when fitting {\it Chandra} X-ray spectra. We created 100 mock effective area curves using the 5--15\% {\it Suzaku} effective area calibration uncertainties reported by \cite{marshall21}, and then used these 100 mock curves to fit 100 spectra. The reported parameter uncertainties from a single Markov Chain Monte Carlo (MCMC) run reflect the statistical uncertainty from fitting ($\sigma_{i, {\rm stat}}$), and the spread in parameter values between all 100 runs represents the telescope calibration uncertainty ($\sigma_{\rm cal}$). The final uncertainty on best-fit spectral parameters is
\begin{equation}\sigma_{\rm param}^2 = \overline{\sigma_{i, {\rm stat}}}^2 + \sigma_{\rm cal}^2\end{equation}

Similar to our analysis in HA23, we then accounted for the fact that obtaining 
total masses from different ejecta components requires assumptions about the unknown emitting volumes (i.e., filling factors) of each component. We calculate mass ratios four times, each time making a different---but equally valid from a physics perspective---assumption about the relations between the component filling factors: 
\begin{enumerate}
\item Each ejecta component has equal emitting volumes.
\item Electron pressure equilibrium P${_e}$=$n_ekT_e$ is constant between ejecta components.
\item Plasmas are linked to specific annuli of the SNR. The Fe-dominated plasma fully fills an inner shell extending from the reverse-shock radius (R$_{\rm RS}$) to the radius where the ejecta becomes IME-dominated (R$_{\rm IME}$). The two IME-dominated plasmas fill a shell from R$_{\rm IME}$ to the contact discontinuity R$_{\rm CD}$. 
\item As above, but we additionally assume that the IME-dominated shells are in electron pressure equilibrium with the Fe-dominated shell. 
\end{enumerate}
The equations associated with the above assumptions are presented in Section~2.4 and Appendix~B of HA23. For assumptions 3.) and 4.), we used the multitude of radius estimates from the literature: R$_{\rm RS}=$ 0.48--0.73R$_{\rm FS}$, R$_{\rm IME}\approx$ 0.75R$_{\rm FS}$, and $R_{\rm CD}=$ 0.84--0.93R$_{\rm FS}$ \citep{warren05,yamaguchi14,katsuda15,millard22,uchida24}. We took the average of the reported values and propagated the spreads as uncertainties throughout our calculations.

\begin{figure}
\begin{center}
\includegraphics[width=\columnwidth]{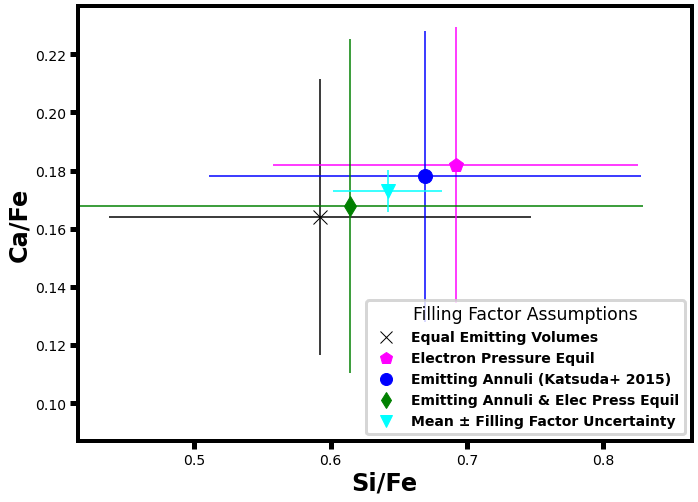} 
\end{center}
\vspace{-5mm}
\caption{\footnotesize{Sample ejecta mass ratios calculated using different filling factor assumptions (non-cyan data points); the cyan triangle is the average of these four estimates. The error bars on the non-cyan points reflect combined photon $+$ telescope calibration uncertainties. The error bars on the cyan data point reflect the spread of estimated mass ratios using the four different filling factor assumptions. }}
\label{fig:diff_methods}
\end{figure}

Figure~\ref{fig:diff_methods} shows a sample of mass ratios calculated assuming different filling factor relations. The error bars on the cyan data point reflect the spread due to the four assumptions, which is an additional source of uncertainty to include in our final mass ratio estimates. To quantify this uncertainty, we used Multiple Imputation \citep{rubin87,schafer97}, a technique designed to handle missing data. The total uncertainty is given by 
\begin{equation}\sigma_{\rm tot}^2 = \overline{\sigma_{\rm fit}^2}+\bigg(1+\frac{1}{M} \bigg)\sigma_{\rm vol}^2\end{equation}
\\
where M is the number of imputations (four), $\overline{\sigma_{\rm fit}}$ is the average of the four mass ratio uncertainties obtained via propagating best-fit parameter uncertainties, and $\sigma_{\rm vol}$ is the standard deviation of those four mass ratio estimates. Finally, we calculated the number of degrees of freedom :
\begin{equation}{\rm degrees\ of\ freedom} = (M-1)\bigg(1+ \frac{M \sigma_{\rm fit}^2}{(M+1)\sigma_{\rm vol}^2}\bigg)^2\end{equation}
and used a t-distribution to find the appropriate 68.3\% confidence interval for our mass ratios.

\section{Results}
\label{sec:results}
Our final mass ratios and their 1$\sigma$\ uncertainties are reported in Table~\ref{table:MRs}. This table also includes the individual 1$\sigma$ uncertainties from both our filling factor assumptions and from MCMC Xspec fitting.

We note that our spectral fit reflects only the bright, shocked X-ray emitting material, not unshocked material inside of the reverse-shock radius. \cite{katsuda15} estimated a shocked-ejecta percentage of $\sim$86\% based on comparisons of their observed masses to the total mass predictions of Ia models and assuming that the vast majority of the interior ejecta were IGEs \citep{badenes06, katsuda15}. Using this value, the true IME/Fe total mass ratios would be $\sim$20\% smaller. We investigate the effects of accounting for varying amounts of unshocked ejecta in more detail in Section~\ref{sec:simuls}.

Additionally, we tied the Ni abundances in our IME-dominated components to the Fe abundances, as leaving Ni to vary freely resulted in it being unconstrained. This means that our estimated Ni/Fe mass ratio is slightly inaccurate, weighted toward the (Ni/Fe)$_\odot$ mass ratio of 0.044. However, as the vast majority ($\sim$85\%) of Fe \& Ni are captured by the Fe-dominated component, the difference between our estimate and the true Ni/Fe mass ratio in Tycho's SNR should be small. As an extreme test case, twice or half as much Ni in the IME-dominated components result in calculated Ni/Fe mass ratios of 0.03 and 0.022, respectively---values well within our Ni/Fe mass ratio confidence interval.

\begin{deluxetable}{lccrc}[!t]
\tablecolumns{4}
\tablewidth{0pt} 
\tablecaption{Calculated Mass Ratios in Tycho's SNR\label{table:MRs}} 
\tablehead{ \colhead{Mass Ratio} & \colhead{Final Value $\pm$ 1$\sigma$} &\colhead{$\sigma_{\rm vol}$\tablenotemark{a}} &\colhead{$\sigma_{\rm fit}$\tablenotemark{b}} }
\startdata
Si/Fe & 0.64 $\pm$ 0.15 & 0.054   & 0.139    \\
Si/S  &  1.47 $\pm$ 0.08 & 0     & 0.08 \\
S/Fe  & 0.44 $\pm$ 0.11 & 0.036  & 0.097 \\
Ar/Fe & 0.14 $\pm$ 0.04 & 0.006   & 0.040   \\
Ar/S  &  0.32 $\pm$ 0.10 & 0.014   & 0.093 \\
Ca/Fe &  0.17 $\pm$ 0.05 & 0.010   & 0.045   \\
Ca/S  & 0.40 $\pm$ 0.09 & 0.013   & 0.088 \\
Cr/Fe & 0.025 $\pm$ 0.002& 0.0004 & 0.0023\\
Mn/Fe & 0.010 $\pm$ 0.002 & 0.0002 & 0.0023 \\
Ni/Fe\tablenotemark{c} & 0.025 $\pm$ 0.009 & 0.0003 & 0.0091  \\ \hline
\enddata
\tablenotetext{a}{Filling factor uncertainties}
\tablenotetext{b}{Combined photon $+$ telescope calibration uncertainties.}
\tablenotetext{c}{The Ni abundance was tied to Fe for the two IME-dominated components (ej1 and ej2).
}
\vspace{-8mm}
\end{deluxetable}

\begin{deluxetable*}{lccccccc}[!h]
\tablecolumns{6}
\tablewidth{0pt} 
\tablecaption{Mass Ratio Comparisons \label{table:prevMRs}} 
\tablehead{ \colhead{Element } & \colhead{Our} & \colhead{Katsuda15}  &\colhead{{Mart{\'\i}nez-Rodr{\'\i}guez}17}  &\colhead{Badenes08} & {Yamaguchi15}\\
\colhead{Ratio} &\colhead{Values\tablenotemark{a}}
&\colhead{Values\tablenotemark{b,c}} & \colhead{Values\tablenotemark{a,b}}  &\colhead{Values\tablenotemark{a,e}}
&\colhead{Values\tablenotemark{f,g}}}
\startdata
Si/Fe & 0.64 $\pm$ 0.25 
& 2.12$^{+1.01}_{-0.72}$ &  & & \\
Si/S &  1.47 $\pm$ 0.13 
& 1.24$^{+1.22}_{-1.18}$ & &  & \\
S/Fe & 0.44 $\pm$ 0.18 
& 1.71$^{+0.86}_{-0.62}$ & &  & \\
Ar/Fe &0.14 $\pm$ 0.07 
& 0.54$^{+0.23}_{-0.17}$ & &  & \\
Ar/S & 0.32 $\pm$ 0.16 
& 0.31$^{+0.30}_{-0.29}$  & 0.218$^{+0.022}_{-0.010}$&  & \\
Ca/Fe &  0.17 $\pm$ 0.08 
& 0.88$^{+0.35}_{-0.25}$  & & & \\
Ca/S & 0.40 $\pm$ 0.15 
& 0.52$^{+0.50}_{-0.46}$ & 0.252$^{+0.025}_{-0.011}$&  & \\
Cr/Fe & 0.025 $\pm$ 0.004 
& & 0.016$^{+0.018}_{-0.005}$& & \\
Mn/Fe & 0.010 $\pm$ 0.004 
& & & & 0.012$^{+0.009}_{-0.004}$  \\
Mn/Cr & 0.41 $\pm$ 0.16 
& & & 0.74$\pm$ 0.47 \\
Ni/Fe & 0.025 $\pm$ 0.015 
& 0.157\tablenotemark{d} & & & 0.023$^{+0.02}_{-0.01}$ \\ 
\enddata
\tablenotetext{a}{90\% Confidence Range}
\tablenotetext{b}{Models using the AtomDB v3.0.9 Fe-K ionization emissivities, which overestimates IME/Fe mass ratios by $\sim$15\% and IGE/Fe mass ratios by $\sim$30\% for plasmas with low ($\lesssim 5 \times 10^{10}$ cm$^{-3}$ s) ionization timescales.}
\tablenotetext{c}{Uncertainties artificially increased from the too-small statistical uncertainties obtained from Xspec fitting.}
\tablenotetext{d}{Ni abundance was tied to Fe abundance} 
\tablenotetext{e}{Calculated using flux and emissivity ratios of \cite{tamagawa09} instead of best-fit abundances, using the equation M$_{\rm Mn}$/M$_{\rm Cr}$=1.057(F$_{\rm Mn}$/F$_{\rm Cr}$)/(E$_{\rm Mn}$/E$_{\rm Cr}$) where F and E are line fluxes and emissivities, respectively.}
\tablenotetext{f}{Calculated using flux and emissivity ratios of \cite{yamaguchi14}, who only analyzed the NW rim of Tycho's SNR.}
\tablenotetext{g}{68\% Confidence Range}
\end{deluxetable*}

\section{Previous Observational Studies}
\label{sec:pastwork}

\label{subsec:prevpapers}
We compare our work to the previous observational studies of \cite{katsuda15}, \cite{mr17}, \cite{badenes08}, and \cite{yamaguchi15}, each of whom used different observations, regions, and techniques to estimate mass ratios in Tycho's SNR. Table~\ref{table:prevMRs} presents the ejecta mass ratios calculated in these works alongside our estimates.

Our choice of model was heavily inspired from the comprehensive work of \cite{katsuda15}, who analyzed the global X-ray spectrum of Tycho's SNR using a multicomponent model. Our spectral model leaves Cr, Mn, and Ni as free parameters so that we could obtain constraints on their abundances.
Although \cite{katsuda15} find that the IGE/IME ratio of Tycho's SNR favors either a subluminous or normally luminous SN Ia, we obtain a higher IGE-to-IME ratio that more strongly favors a SN Ia of moderate luminosity. We note that although their IME/Fe mass ratios are $\sim$2$\sigma$ higher than ours, enforcing electron pressure equilibrium on their models resulted in mass ratios consistent with ours, supporting the moderate-luminosity Type Ia origin. This conclusion matches with the results of other studies \citep{ruizlapuente04,badenes08,krause08}.

\cite{mr17} performed a similar global analysis, but fit separate models to the 2--6~keV (Si, S, Ar, and Ca K$\alpha$ emission lines) and 5--8~keV (Cr, Mn, Fe, and Ni lines) emission. Given that our models required significant Ar and Ca contribution in the hottest, Fe-dominated component, it is unsurprising that our mass ratio estimates differed from theirs. The results of \cite{mr17} favored high detonation densities (for SD origins), low primary masses (for DD origins), and $\sim$1.8Z$_\odot$ progenitors. Compared to the same explosion models, our results favor subsolar metallicity progenitors and lower detonation densities.

\cite{badenes08} converted the Mn/Cr flux ratio measured by \cite{tamagawa09} to a mass ratio, and then used the scaling relation M$_{\rm Mn}$/M$_{\rm Cr} = 5.3 \times Z^{0.65}$ to convert their Mn/Cr mass ratio to a metallicity of 0.048$^{+0.051}_{-0.036} \approx$ 3$\pm$2Z$_\odot$. They suggested that, due to the combination of its high metallicity and large Galactocentric radius, Tycho's SNR likely comes from the ``prompt'' channel of Type Ia SNe. Using our Mn/Cr mass estimate in this scaling relation, we obtain Z$\approx$ 0.019$\pm$0.007 $\approx$ (1$\pm$ 0.37) Z$_\odot$. This metallicity matches better with the average for Tycho's location \citep{nordstrom04} and allows for an older progenitor.

\cite{yamaguchi15} converted IGE flux ratios measured by \cite{yamaguchi14}---extracted from a large Fe-dominated NW region of Tycho's SNR---to mass ratios. Their results favor a metallicity of 1.8Z$_\odot$, and are consistent with either a sub-M$_{\rm Ch}$ detonation with a progenitor WD mass of $\sim$0.93M$_\odot$ or a near-M$_{\rm Ch}$ DDT where all of the emission comes from incomplete Si-burning and nuclear statistical equilibrium (NSE) rather than  neutron-rich NSE (n-NSE). Our estimates are consistent with theirs.

\section{Type Ia Simulations and Origin Scenarios}
\label{sec:simuls}
In this section, we compare our mass ratio estimates to the results of various near- and sub-M$_{\rm Ch}$ Type Ia simulations. In particular, as mentioned in Section~\ref{sec:results}, there is an unknown amount of unshocked (and thus not emitting in X-rays) ejecta present in the central regions of Tycho's SNR. Thus, we extract nucleosynthesis data that excludes differing amounts of the innermost ejecta and compare each set of derived mass ratios to our estimates.

\begin{figure*}
\begin{center}
Type Ia Near-M$_{\rm Ch}$ Nucleosynthesis Results \\
\includegraphics[width=\textwidth]{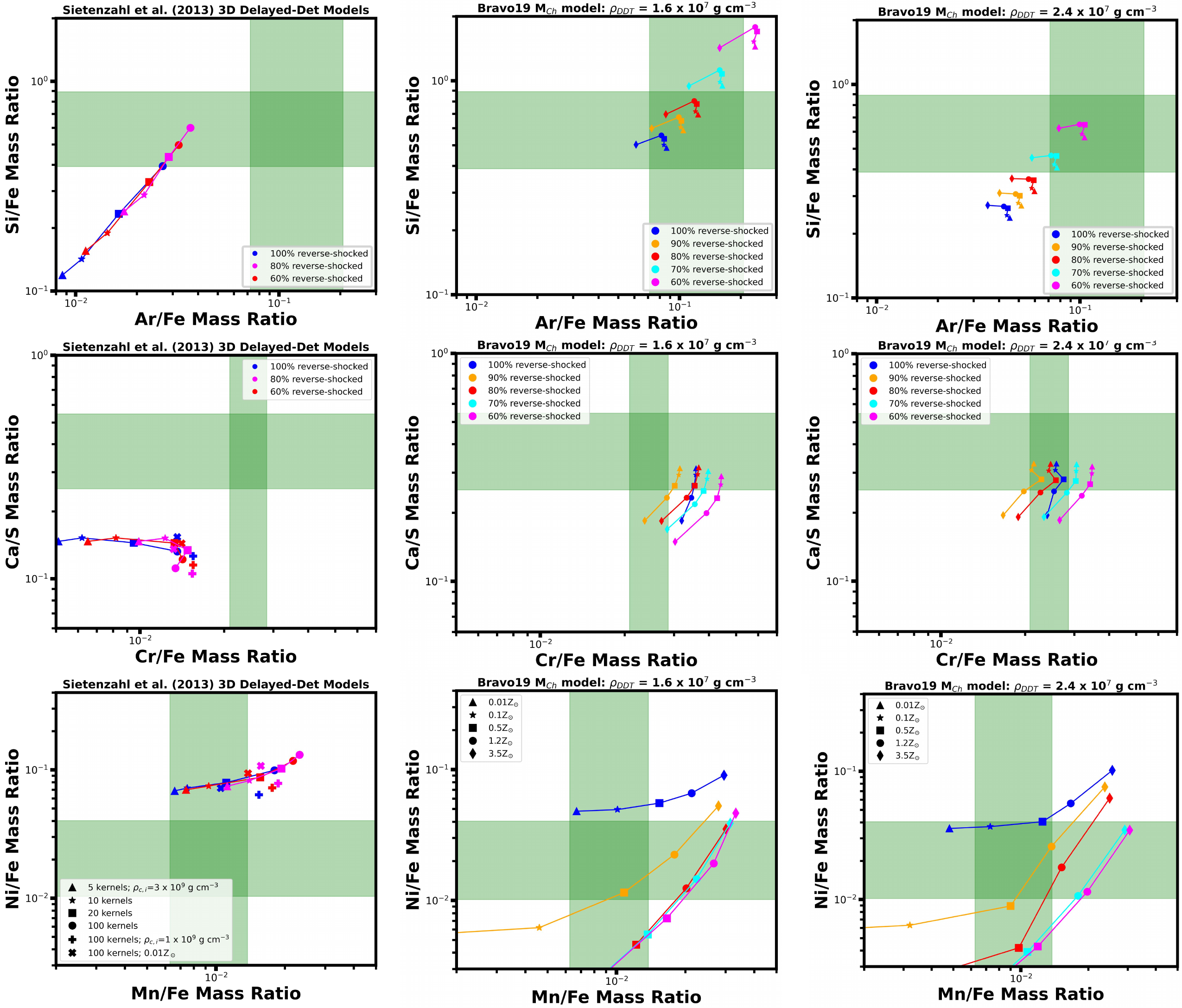} \\
\includegraphics[width=0.58\textwidth]{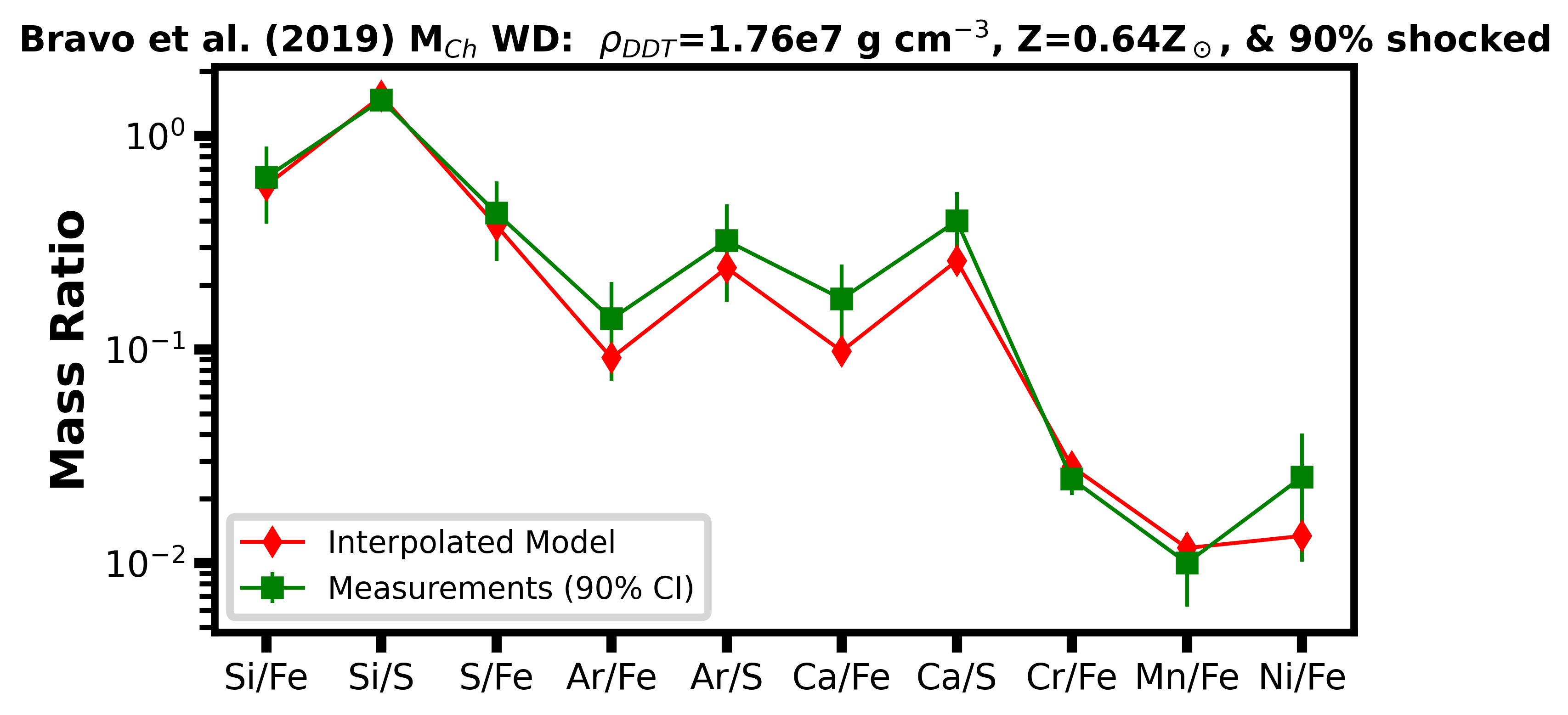}
\end{center}
\vspace{-5mm}
\caption{\footnotesize{Our estimated mass ratios for ejecta in Tycho's SNR (green bars represent 90\% confidence intervals) compared to the results of various near-M$_{\rm Ch}$ Delayed Detonation nucleosynthesis models. Each column of three plots corresponds to the same simulation (with labels split between the three plots), and each row shows different mass ratios. The bottom plot shows results interpolated from the multiple models of \cite{bravo19} that best matches (reduced-$\chi^2$ = 1.76) our mass ratio estimates.}}
\label{fig:NearCh_simulations}
\end{figure*}

\subsection{Near-M$_{\rm Ch}$ Explosion Models}

In Figure~\ref{fig:NearCh_simulations}, we present the 90\% confidence intervals of our mass ratio estimates compared to the nucleosynthesis yields of various near-M$_{\rm Ch}$ simulations. Each column shows the predictions of a specific model, and each row shows a different set of mass ratios. Within each plot, the colors represent differing amounts of exterior X-ray-emitting ejecta.

The left column of Figure~\ref{fig:NearCh_simulations} shows the nucleosynthesis yields for 3D near-M$_{\rm Ch}$ DDT models of \cite{seitenzahl13}. 
Compared to our estimates, their models underproduce Si/Fe, Ar/Fe, Ca/S, and Cr/Fe, and overproduce Ni/Fe. Our results are inconsistent with the predictions of their simulations, and this finding remains true even when accounting for different amounts of unshocked ejecta. We also compared our results to the classic 1D W7 (pure deflagration) and WDD2 (detonation-to-detonation) models, as well as the 2D deflagration-to-detonation simulations of \cite{leung18}. All of these simulations failed to match our observed mass ratios for Tycho's SNR in similar manners to those of \cite{seitenzahl13}.

The middle and right rows of Figure~\ref{fig:NearCh_simulations} show nucleosynthesis results from \cite{bravo19}, who used a 90\% attenuated $^{12}$C$+^{16}$O reaction rate in their SN simulations. Their simulations produced Ca/S mass ratios that match much better with our estimates compared to simulations that use the standard reaction rate. It should be noted that \cite{bravo19} manually adjusted the reaction rate physics in order to reproduce the higher Ca/S mass ratios observed in a few Type Ia SNRs. It is promising that our results---involving fitting to a wider X-ray bandpass and to global SNR emission---still remain well-described by their simulations.

\begin{figure*}
\begin{center}
Type Ia Sub-M$_{\rm Ch}$ Nucleosynthesis Results \\
\includegraphics[width=\textwidth]{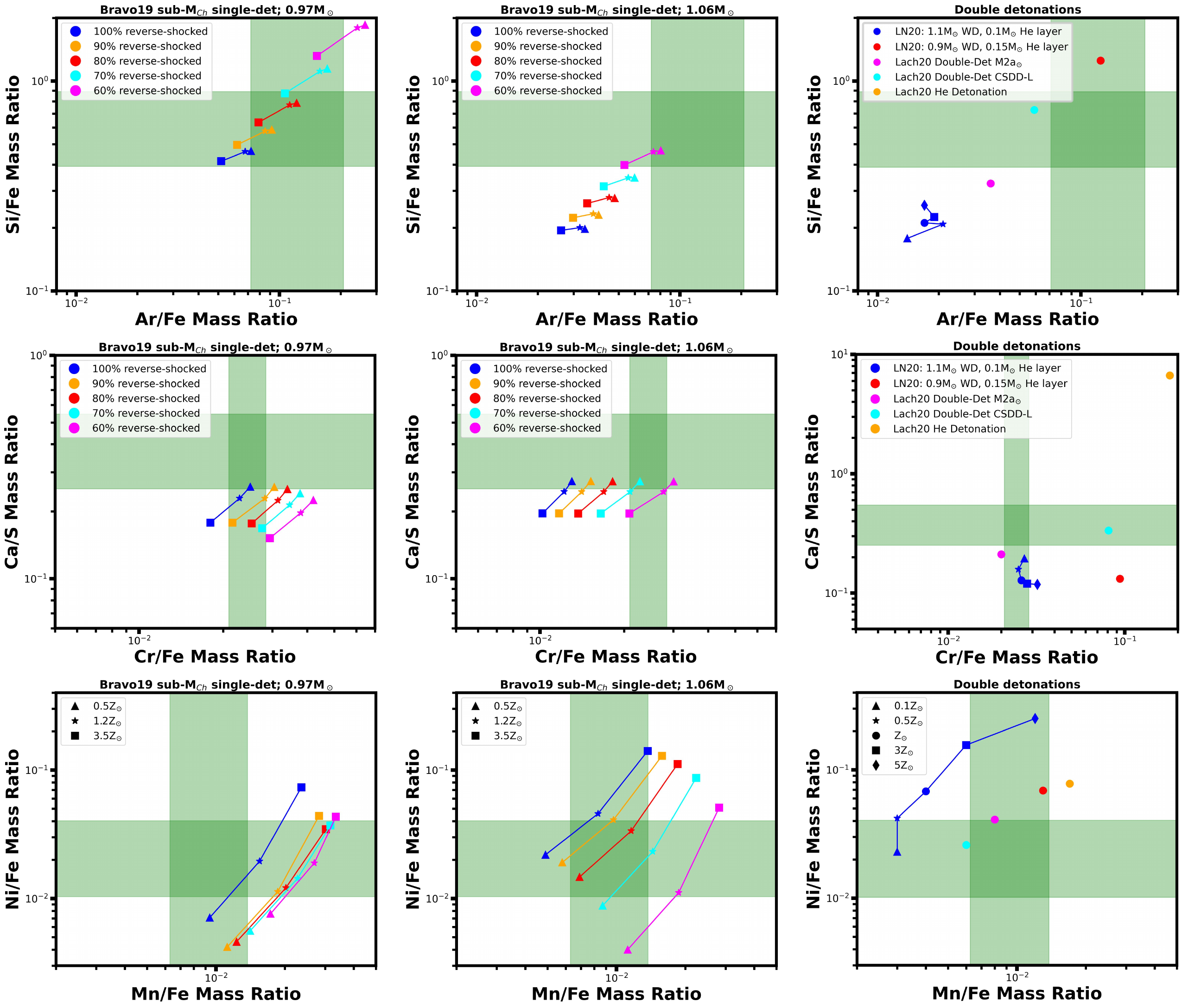} \\
\includegraphics[width=0.58\textwidth]{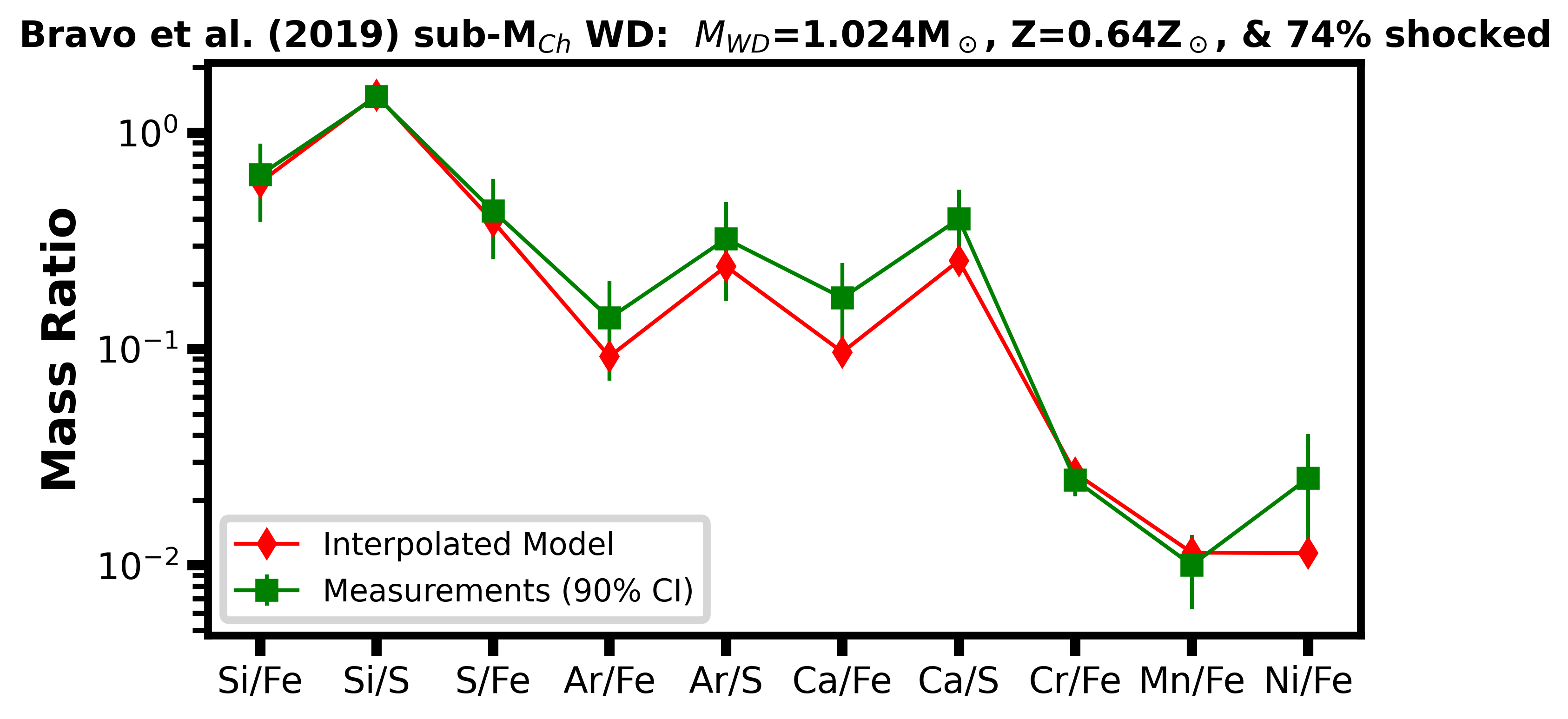}
\end{center}
\vspace{-5mm}
\caption{\footnotesize{Same as Figure~\ref{fig:NearCh_simulations}, but showing the simulated nucleosynthesis results from \cite{bravo19} single-detonation and the \cite{leung20,lach20} double-detonation sub-M$_{\rm Ch}$ simulations. Each column corresponds to the same simulation (with labels split between the three plots), and each row shows different mass ratios. 
The bottom plots show results interpolated from the multiple models of \cite{bravo19} that best matches (reduced-$\chi^2$ = 1.50) our mass ratio estimates.}}
\label{fig:SubCh_simulations1}
\end{figure*}

We show two sets of M$_{\rm Ch}$ simulations results from \cite{bravo19}: those using deflagration-to-detonation transition densities of 1.6 (middle) and 2.4 (right) $\times 10^7$ g cm$^{-3}$. We choose these densities as their results appear to bracket our estimated mass ratios; a model using an interpolation between the two densities should match well with our results. We assumed linear interpolation with 10 equally spaced steps between the variables: density, metallicity, and shocked-ejecta percentage. The final plot in Figure~\ref{fig:NearCh_simulations} shows the combination of parameters that produce mass ratios with the lowest reduced-$\chi^2$ value when compared to our estimates: a value of 1.76 ($\chi^2$=8.80). The $\chi^2$ value stays below the 90\% critical value of 9.23 using 5 degrees of freedom (reflecting the 8 elements whose abundances we measured, minus the 3 simulation variables) for parameter ranges of $\rho$=1.65--1.9 $\times 10^7$ g cm$^{-3}$, Z=0.47--0.83Z$_\odot$, and with 89--91\% of the ejecta reverse-shock heated. When calculating the reduced-$\chi^2$, we didn't include any IME/S mass ratios as they are fully degenerate with other mass ratios.

\begin{figure*}
\begin{center}
Type Ia Sub-M$_{\rm Ch}$ Nucleosynthesis Results \\
\includegraphics[width=\textwidth]{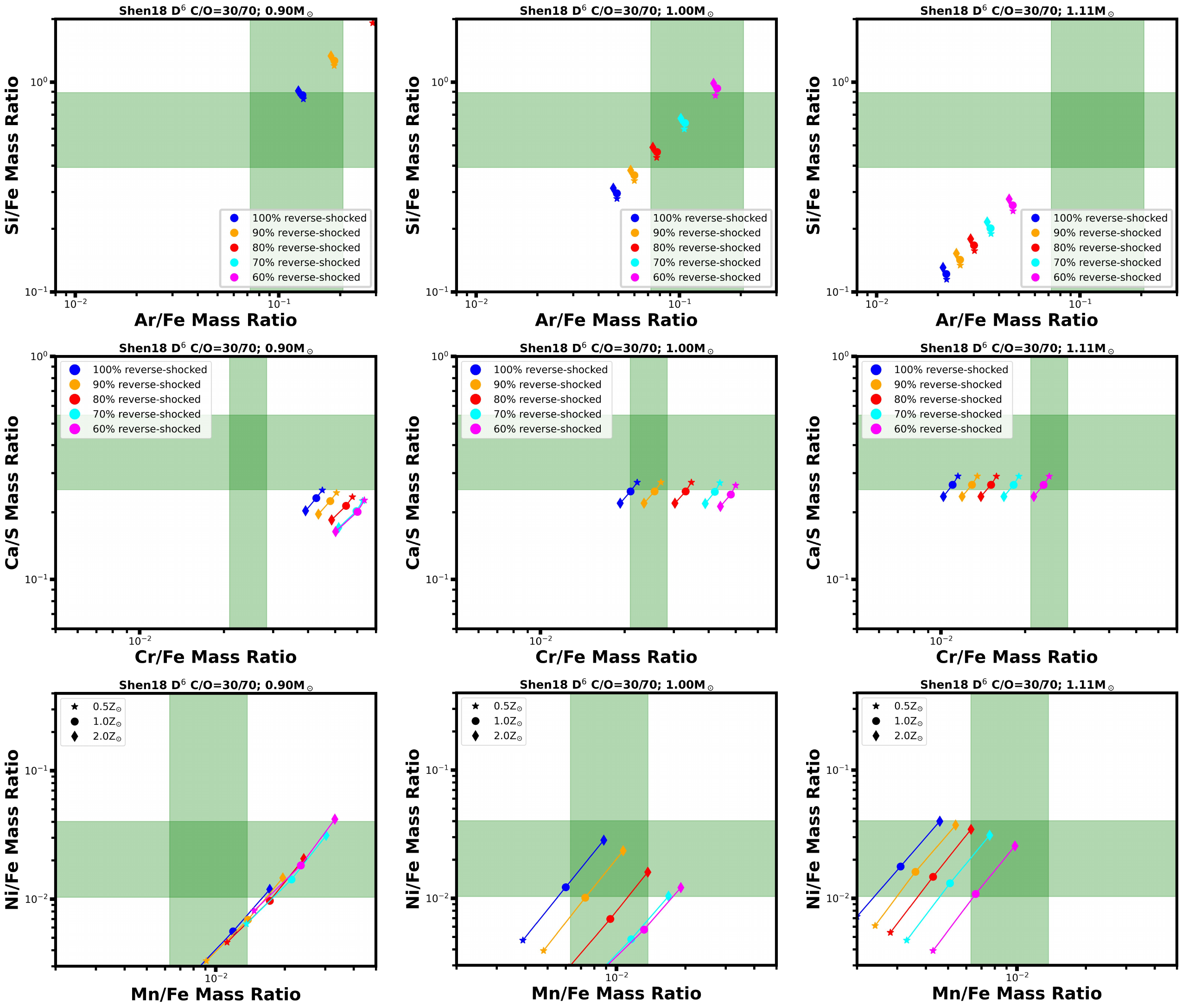}\\
\includegraphics[width=0.51\textwidth]{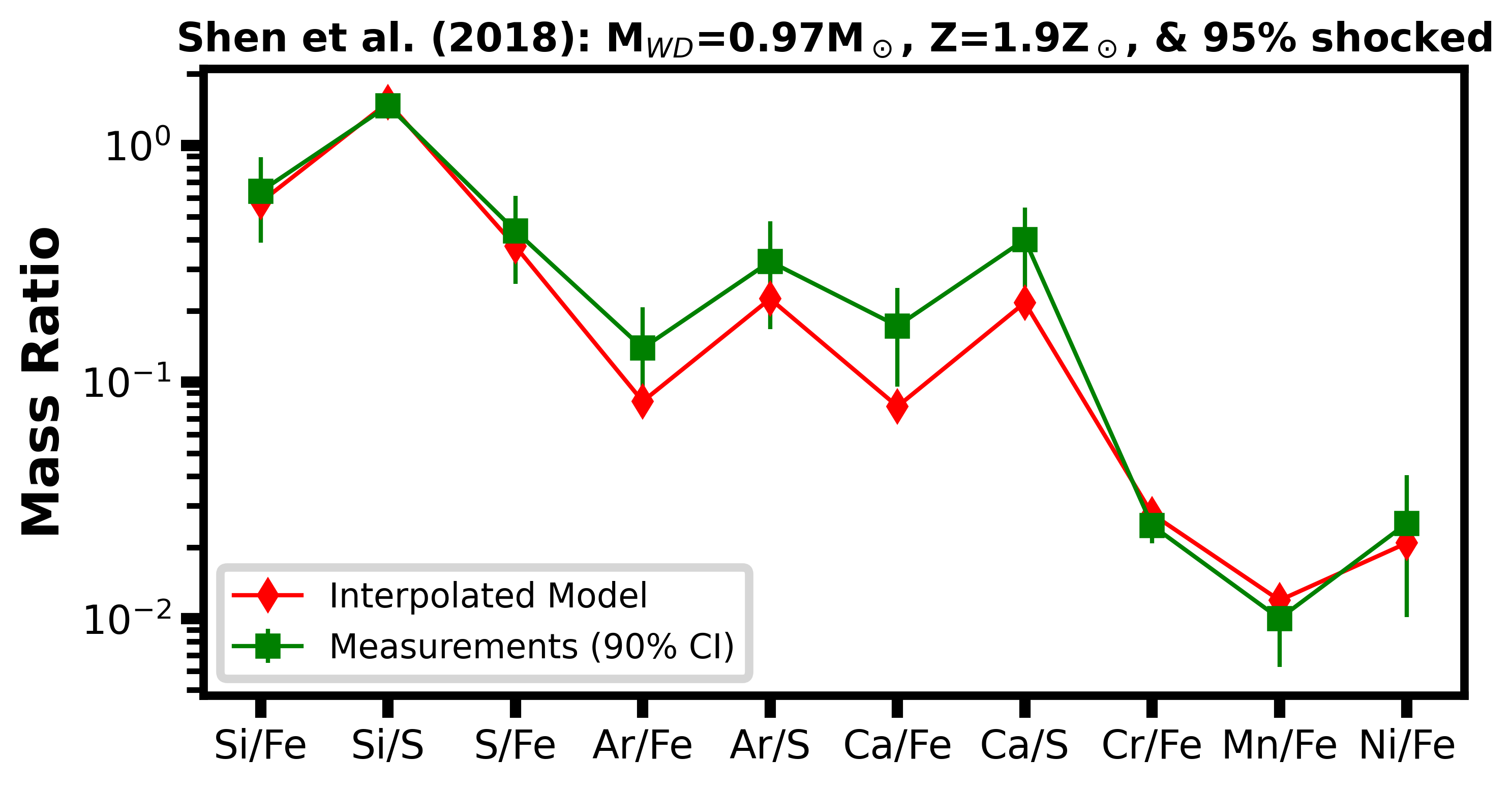}
\\
\end{center}
\vspace{-5mm}
\caption{\footnotesize{Same as Figure~\ref{fig:NearCh_simulations}, but showing the simulated nucleosynthesis results from \cite{shen18a} sub-M$_{\rm Ch}$, D$^6$ nucleosynthesis models. Each column of three plots corresponds to the same simulation (with labels split between the three plots), and each row shows different mass ratios. The bottom plot show results interpolated from the multiple models of \cite{shen18a} that best matches (reduced-$\chi^2$ = 1.77) our mass ratio estimates.}}
\label{fig:SubCh_simulations2}
\end{figure*}

\subsection{Sub-M$_{\rm Ch}$ Explosion Models}
In Figure~\ref{fig:SubCh_simulations1} and Figure~\ref{fig:SubCh_simulations2}, we present our mass ratio estimates compared to the nucleosynthesis yields of various sub-M$_{\rm Ch}$ simulations. These models include a mix of single detonations, stable double detonations (DDet), and dynamically unstable double detonations (D$^6$).

In the left two columns of Figure~\ref{fig:SubCh_simulations1}, we present mass ratios from the sub-M$_{\rm Ch}$ models of \cite{bravo19} using the 90\% attenuated $^{12}$C$+^{16}$O reaction rate. The format of the figure is the same as previous ones; each column represents a specific model (in this case, the progenitor WD mass), and each row shows different mass ratios. Similar to the near-M$_{\rm Ch}$ models of \cite{bravo19}, no specific model reported by them matches with the 90\% confidence intervals of our estimates. However, an interpolation between published models does.

The bottom row of Figure~\ref{fig:SubCh_simulations1} presents the linearly interpolated results from \cite{bravo19} that best match our mass ratio estimates: a progenitor mass of 1.024M$_\odot$, a metallicity of 0.64Z$_\odot$, ejecta that has been 74\% shocked (by mass), and a reduced-$\chi^2$ value of 1.50 ($\chi^2\approx7.52$). The $\chi^2$ value remains below a 90\% critical value of 9.24 for parameter ranges of: M=0.997--1.042M$_\odot$, Z=0.32--0.95Z$_\odot$, and 67--85\% propagation of the reverse shock through the ejecta.

In the right column of Figure~\ref{fig:SubCh_simulations1} we present the 2D DDet models of \cite{lach20} and \cite{leung20}. Their models are inconsistent with our mass ratio estimates for Tycho's SNR, generally overproducing Ni/Fe and Cr/Fe while underproducing Si/Fe and Ca/S. Although we could not obtain precise Lagrange mass coordinate data for these models, other simulations exhibited trends of higher IGE/Fe ratios with greater amounts of reverse shocked ejecta; this effect would exacerbate the inconsistencies between our estimates and the predictions of \cite{lach20,leung20}. Thus, we can exclude a DDet origin scenario for Tycho's SNR.

In Figure~\ref{fig:SubCh_simulations2}, we present mass ratios from the 1D, sub-M$_{\rm Ch}$ models of \cite{shen18a}, representing D$^6$ explosions.  They performed simulations using both the standard C$+$O reaction rate and the 90\% attenuated rate---we only use data from the latter simulations, as the former ones completely fail to match the observed Ar- and Ca-to-S mass ratios. Additionally, only their models with initial WD C/O ratios of 30:70 produced estimates consistent with our results---thus, we don't include any results using C/O ratios of 50:50. 

We note that this is the opposite of our analysis of Kepler's SNR, where simulations with progenitor C/O ratios of 50:50 were favored. However, the \cite{shen18a} simulations were statistically inconsistent with our Kepler mass ratio estimates at the 95\% confidence level. 
Although the trends in the plots suggest that models with Z$>$2Z$_\odot$ might be consistent, \cite{shen18a} did not investigate this parameter space. Thus, it is difficult to draw any specific conclusions about the progenitor C/O ratios in Kepler's vs. Tycho's SNRs.

The bottom row of Figure~\ref{fig:SubCh_simulations2} presents the best-matching interpolated results from \cite{shen18a}: a progenitor mass of 0.97M$_\odot$, a metallicity of 1.9Z$_\odot$, ejecta that has been 95\% shocked by mass. The corresponding a reduced-$\chi^2$ value of 1.77 ($\chi^2\approx8.83$) falls below the 90\% critical value of 9.23 for parameter ranges of: M = 0.96--0.98M$_\odot$, Z$\gtrsim$1.6Z$_\odot$, and 90--98\% shocked ejecta.

\section{Conclusions}
\label{sec:conc}
In this paper, we have performed comprehensive spectral analysis of the 0.6--8.0~keV X-ray emission from Tycho's SNR. We fit the global emission with a multicomponent plasma model, accounting for swept-up ISM/CSM and nonthermal synchrotron emission. We used MCMC fitting that accounted for the 5--15\% effective area calibration uncertainty in {\it Suzaku} XIS detectors in order to obtain robust best-fit parameters and uncertainties. Due to the incredibly high amount of photons detected from this bright SNR---particularly between 0.6 and 3.0~keV---the telescope calibration uncertainty dominated over the statistical photon uncertainty. 

Additionally, we accounted for the unknown filling factors associated with each component of our multicomponent spectral model by making reasonable physics-based assumptions about possible relations between the ejecta components. In contrast to our findings for Kepler's SNR---where these unknowns could lead to uncertainties of $\sim$30\%---these uncertainties were $\lesssim$10\% for Tycho's SNR and were universally smaller than the statistical fit $+$ systematic telescope effective area uncertainties.

We conclude the following about Tycho's progenitor:
\begin{enumerate}
    \item Our IME/S mass ratio estimates were only reproduced by simulations that used a 90\% attenuation of the standard $^{12}$C$+^{16}$O reaction rate (e.g., \citealt{mr17,shen18a,bravo19}): similar to our findings for Kepler's SNR. 
    \item Our findings most strongly favor a Type Ia explosion of normal luminosity where the vast majority ($\sim$75--95\%) of the ejecta has been shock heated by the reverse shock at this time.
    \item The nucleosynthesis predictions of dynamically stable double detonations are inconsistent with our mass ratio estimates. 
    \item Both near- and sub-M$_{\rm Ch}$ 1D progenitor models from \cite{bravo19} are consistent with our mass ratio estimates. The former requires a relatively low deflagration-to-detonation transition density ($\sim$1.6 $\times 10^7$ g cm$^{-3}$) and the latter requires a 1.0--1.05M$_\odot$ progenitor mass. Both models support progenitor metallicities of Z$\approx$0.64Z$_\odot$ and use the 90\% attenuated $^{12}$C$+^{16}$O reaction rate. 
    \item The 1D, D$^6$ models from \cite{shen18a} with a progenitor mass of $\sim$1M$_\odot$ and a metallicity of 1.9Z$_\odot$ also are consistent with our results. These models produce Ni/Fe mass ratios much closer to our estimates, along with metallicities much closer to values reported in past papers on Tycho's SNR. The largest discrepancies are that the Ar/S \& Ca/S mass ratios from these models don't match our estimates. 
\end{enumerate}
We recommend that 2D and 3D models using the 90\% attenuated $^{12}$C$+^{16}$O reaction rate and parameter values matching our interpolated best-matches should be investigated to confirm their consistency with our estimated ejecta mass ratios in Tycho's SNR. We also recommend that further lab studies to refine the cross section of the $^{12}$C$+^{16}$O reaction rate should be performed.

Alternatively, it has been suggested that this 90\% C$+$O reaction rate attenuation can be reproduced by smaller variations in multiple reactions involving $^{12}$C and $^{16}$O \citep{bravo19} and/or the efficiency of neutronization or carbon simmering prior to explosion \citep{chamulak08}. Both of these scenarios should be further investigated through lab studies and simulations that attempt to reproduce observed properties of Tycho's SNR and SNRs of all types.

Observations with {\it XRISM} and future observatories such as {\it AXIS} and {\it NewAthena} will allow analysis on finer spatial and spectral scales, enabling astronomers to disentangle line components, measure odd-Z elements, create detailed 3D ejecta maps, and obtain better constraints on the ionization states of different ejecta elements in supernova remnants.

\acknowledgments
\noindent
\textbf{Acknowledgments}

We thank Dr. Adam Foster for providing updated AtomDB Fe emission line data and Dr. Keith Arnaud for providing advice on obtaining mass ratios for non-hydrogen-dominated plasmas. We thank Drs. Eduardo Bravo, Carles Badenes, Ivo Seitenzahl, and Ken Shen for providing Lagrange Mass Coordinate nucleosynthesis tables from their simulations' results. We thank the referee for their insightful comments.

Tyler Holland-Ashford’s research was supported by an appointment to the NASA Postdoctoral Program at the NASA Goddard Space Flight Center, administered by Oak Ridge Associated Universities under contract with NASA. Dr. Pat Slane acknowledges support from NASA Contract NAS8-03060.

This research made use of the analysis software: HEASoft (6.31, \small http://heasarc.gsfc.nasa.gov/ftools, \normalsize ascl:1408.004), XSPEC (v12.13.0; \citealt{arnaud96}), AtomDB (v3.0.10 \citealt{smith01,foster12}), PyAtomDB (\small https://atomdb.readthedocs.io/en/master/), \normalsize and CALDB (\small https://heasarc.gsfc.nasa.gov/FTP/caldb, \normalsize XIS 20160607). This research made use of data obtained from the {\it Suzaku} satellite, a collaborative mission between the space agencies of Japan (JAXA) and the USA (NASA). The computations in this paper were conducted on the Smithsonian High Performance Cluster (SI/HPC), Smithsonian Institution (https://doi.org/10.25572/SIHPC). This work made use of the Heidelberg Supernova Model Archive (HESMA; https://hesma.h-its.org).

\appendix

\section{MCMC Best Fit Model}
\label{appen:comprehensive}
Table~\ref{table:fitvals} presents the average best-fit parameters from our 100 MCMC spectral fits along with the standard deviations between the 100 fits ($\sigma_{\rm cal}$) and the average statistical uncertainty reported by individual Xspec fits ($\overline{\sigma_{\rm stat}}$).

\begin{deluxetable*}{lccc}[!h]
\tablecolumns{4}
\tablenum{3}
\tablewidth{0pt} 
\tablecaption{Final Model Parameters \label{table:fitvals}} 
\tablehead{ \colhead{Parameter} & \colhead{Value $\pm$ 1$\sigma$ } &\colhead{$\sigma_{\rm {cal}}$\tablenotemark{a}} & \colhead{$\overline{\sigma_{\rm {stat}}}$\tablenotemark{a}} }
\startdata
$N_{\rm{H}}$ (10$^{22}$ cm$^{-2}$)  & 1.351 $\pm$ 0.038 & 0.038 & 0.005 \\ \hline
CSM Component: \texttt{vpshock}\\
$kT_{\rm e}$ (keV)   & 0.225 $\pm$ 0.013 & 0.013 & 0.002  \\
Abundance\tablenotemark{b} (solar) O  & 0.91 $\pm$ 0.05 & 0.05& 0.02\\
Si  & 0.32 $\pm$ 0.16 & 0.12 & 0.10 \\
Ionization Timescale ($10^{11}$ cm$^{-3}$ s)  & 1.99 $\pm$ 0.33 & 0.32 & 0.09 \\
Redshift ($10^{-3}$)  & 2.43 $\pm$  3.92 & 3.91& 0.29 \\
Line Broadening (E/6 keV)  & 0.016 $\pm$ 0.004 & 0.004 & 0.001\\
Normalization	(cm$^{-5}$)   & 8.14 $\pm$ 1.80 & 1.76 & 0.35\\\hline
Ejecta 1 component: \texttt{vpshock}\\
$kT_{\rm e}$ (keV)    & 0.827 $\pm$  0.123 & 0.122 & 0.006 \\
Abundance\tablenotemark{b} (10${^4}$ solar) O & 1.97 $\pm$ 0.93 & 0.92 & 0.10 \\
Ne  & 0.07 $\pm$  0.05 &0.04& 0.02\\
Mg  & 0.33 $\pm$  0.11& 0.10& 0.02 \\
Si  & 10\tablenotemark{c}& --& --\\
S  & 9.01 $\pm$  0.64 & 0.63 & 0.05\\
Ar  & 9.26 $\pm$ 1.36 & 1.34 & 0.09\\
Ca  & 18.48 $\pm$ 3.15 & 3.14 & 0.24\\
Fe  & 0.96 $\pm$ 0.17 & 0.16 & 0.04 \\
Ionization Timescale($10^{11}$ cm$^{-3}$ s) & 1.22 $\pm$  0.49 & 0.48& 0.05\\
Redshift ($10^{-3}$)  & 1.52 $\pm$  1.32 & 1.31 & 0.07 \\
Normalization	(10$^{-6}$ cm$^{-5}$)   & 39.67 $\pm$ 6.43 & 6.39 & 0.71 \\\hline
Ejecta 2 component: \texttt{vpshock}\\
$kT_{\rm e}$ (keV)   & 1.65 $\pm$  0.21 & 0.20 & 0.02 \\
Abundance\tablenotemark{b} (10$^4$ solar) Fe  & 0.38 $\pm$  0.20 & 0.19 & 0.04 \\
Ionization Timescale ($10^{10}$ cm$^{-3}$ s)  & 6.86 $\pm$ 0.87 & 0.86 & 0.01\\
Redshift ($10^{-3}$)   & -1.75 $\pm$ 1.69 & 1.69 & 0.06\\
Line Broadening (E/6 keV)  & 0.0220 $\pm$ 0.0029 & 0.0028 & 0.0004\\
Normalization	(10$^{-6}$ cm$^{-5}$)  & 12.71 $\pm$  4.02 & 4.01 & 0.33\\\hline
Ejecta 3 component: \texttt{vvnei}\\
$kT_{\rm e}$ (keV)   & 4.73 $\pm$  0.70 & 0.69 & 0.11 \\
Abundance\tablenotemark{b} (10${^4}$ solar) 
Ar  & 12.88 $\pm$ 2.79 & 2.76 & 0.42 \\
Ca  & 16.35 $\pm$ 1.74 & 1.69 & 0.41 \\
Cr  & 25.41 $\pm$ 2.06  & 1.82 & 0.96 \\
Mn  & 14.40 $\pm$ 3.21 & 2.60 & 1.88 \\
Fe  & 10\tablenotemark{c} & --& --\\
Ni  & 5.22 $\pm$ 2.35 & 2.26 & 0.65 \\
Ionization Timescale ($10^{9}$ cm$^{-3}$ s)  & 2.88 $\pm$ 0.45& 0.44 & 0.08\\
Redshift ($10^{-3}$)  & -3.00 $\pm$ 0.42& 0.40 & 0.11\\
Line Broadening (E/6 keV)  & 0.62 \tablenotemark{c} & -- & --\\
Normalization	(10$^{-6}$ cm$^{-5}$)   & 5.86 $\pm$ 1.44 & 1.42 & 2.30 \\\hline
Synchrotron component: \texttt{srcut}\\
Radio Spectral Index $\alpha$  & -0.6 \tablenotemark{c} & --& --\\
Break Frequency (10$^{17}$ Hz)  & 2.54 $\pm$ 0.36 & 0.36 & 0.04\\
Normalization (1 GHz flux; Jy)   & 10.04 $\pm$ 1.66 & 1.64 & 0.18 \\
\enddata
\vspace{-1mm}
\tablenotetext{a}{\footnotesize{$\sigma_{\rm {cal}}$ and {$\overline{\sigma_{\rm {stat}}}$} reflect the uncertainties from {\it Suzaku} effective area calibration and Xspec fitting, respectively. }}
\vspace{-1mm}
\tablenotetext{b}{\footnotesize{CSM abundances of S, Ar, Ca, \& Fe were tied to Si. Ejecta 1 and Ejecta 2 element abundances were all linked except for that of Fe. All other unmentioned element abundances were fixed to solar. }}
\vspace{-1mm}
\tablenotetext{c}{\footnotesize{Values frozen during fitting. }}
\end{deluxetable*}

\begin{deluxetable*}{lccc}[!h]
\tablecolumns{4}
\tablenum{3}
\tablewidth{0pt} 
\tablecaption{Final Model Parameters, cont. \label{table:fitvals}} 
\tablehead{ \colhead{Parameter} & \colhead{Value $\pm$ 1$\sigma$ } &\colhead{$\sigma_{\rm {cal}}$\tablenotemark{a}} & \colhead{$\overline{\sigma_{\rm {stat}}}$\tablenotemark{a}} }
\startdata
\hline
Gaussians:\\\hline
Energy$_1$ (keV)  & 0.735 $\pm$ 0.002 & 0.002 & 0.001\\
Line Width$_1$ (keV)  & 0.0017 $\pm$ 0.0017 & 0.0013 & 0.0012\\
Normalization$_1$ (photons cm$^{-2}$ s$^{-1}$)   & 0.424$\pm$ 0.110 & 0.109 & 0.012 \\
Energy$_2$ (keV)  & 1.226 $\pm$ 0.003 & 0.003& 0.001\\
Line Width$_2$ (keV)  & 0.010 $\pm$ 0.010 & 0.009 & 0.003\\
Normalization$_2$ (photons cm$^{-2}$ s$^{-1}$)   & 0.0057 $\pm$ 0.0013 & 0.0013 & 0.0002 \\
\enddata
\vspace{-1mm}
\end{deluxetable*}

\bibliography{Tycho_MR}

\end{document}